# Improved in-situ characterization of electrochemical interfaces using metasurface-driven surface-enhanced infrared absorption spectroscopy

Luca M. Berger [1, ‡], Malo Duportal [2, ‡], Leonardo de Souza Menezes [1,3], Emiliano Cortés [1], Stefan A. Maier [4,5,1], Andreas Tittl [1,*], Katharina Krischer [2,*]

[1] *Faculty of Physics, Ludwig-Maximilian-University Munich, 80539 München, Germany*
[2] *Department of Physics, Technical University of Munich, 85748 Garching, Germany*
[3] *Departamento de Física, Universidade Federal de Pernambuco, 50670-901 Recife-PE, Brazil*
[4] *School of Physics and Astronomy, Monash University, Melbourne, Victoria, Australia*
[5] *Department of Physics, Imperial College London, SW7 2AZ London, United Kingdom*

*\* e-mail: Andreas.Tittl@physik.uni-muenchen.de; krischer@tum.de*

**Abstract**

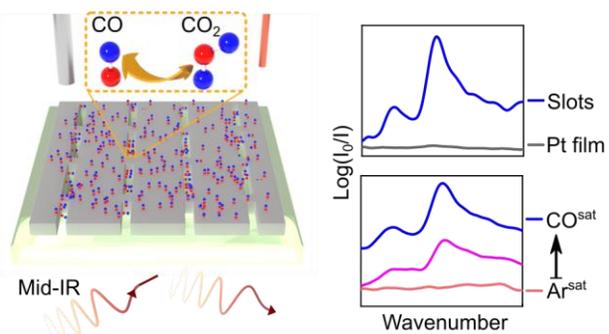

Electrocatalysis plays a crucial role in realizing the transition towards green energy, driving research directions from hydrogen generation to carbon dioxide reduction. Understanding electrochemical reactions is crucial to improve their efficiency and to bridge the gap toward a sustainable zero-carbon future. Surface-enhanced infrared absorption spectroscopy (SEIRAS) is a suitable method for investigating these processes because it can monitor with chemical specificity the mechanisms of the reactions. However, it remains difficult to detect many relevant aspects of electrochemical reactions such as short-lived intermediates. Here, we develop and experimentally realize an integrated nanophotonic-electrochemical SEIRAS platform for the *in situ* investigation of molecular signal traces emerging during electrochemical experiments. Specifically, we implement a platinum nano-slot metasurface featuring strongly enhanced electromagnetic near fields and spectrally target it at the weak vibrational bending mode of adsorbed CO at ~2033 cm$^{-1}$. Crucially, our platinum nano-slot metasurface provides high molecular sensitivity. The resonances can be tuned over a broad range in the mid-infrared spectrum. Compared to conventional unstructured platinum layers, our nanophotonic-electrochemical platform delivers a substantial improvement of the experimentally detected characteristic absorption signals by a factor of 27, enabling the detection of new species with weak signals, fast conversions, or low surface concentrations. By providing a deeper understanding of catalytic reactions, we anticipate our nanophotonic-electrochemical platform to open exciting perspectives for electrochemical SEIRAS, surface-enhanced Raman spectroscopy, and the study of reactions in other fields of chemistry such as photoelectrocatalysis.

**Introduction**

Electrochemical reactions underpin many technologies ubiquitous for a future carbon-zero world such as green-hydrogen generation for long-term sustainable energy storage[1] and $CO_2$ degradation to combat the current trends of climate change[2]. Unfortunately, in general, the monitoring, and therefore understanding, of many electrochemical reactions remains a challenge. In particular, resolving the electrochemical $CO_2$ reduction reaction ($CO_2$RR) with high efficiency, selectivity, and sensitivity remains an issue[3] especially due to the competition with the hydrogen evolution reaction (HER) at high current densities[4]. During the $CO_2$RR to desired carbon products, a compulsory step to the key intermediate CO is still not fully understood and requires further investigation[5].

For the detection and characterization of molecules, optical spectroscopy, mass spectrometry, chromatography, and fluorescence microscopy are often used[6]. Optical spectroscopy methods in particular are highly advantageous because they allow for the retrieval of the spectral fingerprint of molecules *via* the detection of their rotational or vibrational modes. Within optical spectroscopy, two strong methods are Raman and infrared (IR) spectroscopy. The former relies on the inelastic scattering of photons and studies the resulting spectral shift. The latter detects the absorption of light by molecules when the energy of the photons matches the energy of the vibrational modes. The mutual exclusion rule dictates that any mode can be IR active, Raman inactive, and vice-versa but not simultaneously[7]. Therefore, for a given molecular mode either Raman or IR spectroscopy can be used, but not both. Here, the CO vibrational mode under investigation is IR active[8]. Surface-enhanced infrared absorption spectroscopy (SEIRAS) is a derivative technique from conventional infrared spectroscopy based on the enhancement of the local electromagnetic (EM) near fields [9,10]. To increase the sensitivity of either surface enhanced Raman spectroscopy (SERS) or SEIRAS during electrochemical reactions typically a rough metal surface has been chosen to enhance the local electromagnetic (EM) near fields[11]. Rough and highly disordered metallic nm-sized edges coming from perforations and extrusions in the metallic film locally confine and enhance the EM fields. Unfortunately, this approach is random, does not allow for spectral tailoring of plasmonic hotspots, and consequently generates a relatively weak EM near-field enhancement. Even after improvements in the sensitivity of SEIRAS using an attenuated total internal reflection (ATR) geometry[9,10], the characterization of CO adsorption on catalysts is still hampered by weak signal traces[12–14].

We overcome the challenge of detecting weak signal traces by taking inspiration from other fields of nanophotonics. In biomolecular sensing, a plethora of alternatives are used to improve molecular detection using controlled and tuneable EM near-field enhancement via the excitation of resonances through tailored system parameters on the nanoscale. Examples are plasmonic nanoparticles, non-plasmonic nanogap dimers[15], metasurfaces based on plasmonics[16] or exotic phenomena like quasi-bound states in the continuum[17], waveguides[18] or 2D-integrated[19] platforms, among others[20]. Plasmonic-based sensors have become the method of choice in label-free detection of biomolecules. They can be used either as 1) refractive index (RI) sensors or 2) by coupling the resonances to the molecular modes and analysing the perturbation of the intensity either in reflection or transmission[21], termed perturbed intensity sensing here.

In fact, some recent progress has been made to integrate plasmonic structures for refractive index sensing with electrochemistry[21–23]. There are also recent examples of plasmonic structures for perturbed intensity sensing for SERS used to monitor electrochemical reactions[24] or to study the mechanism of an electrocatalytic reaction[25]. Literature of plasmonic imaging provides other examples of electrochemical reactions of single nanoparticles[26], plasmonics-supported and electrochemical monitoring of molecular interactions focused on fluorescence and confocal microscopy[27,28], and plasmon-accelerated electrochemical reactions[29,30]. However, to the best of our knowledge, the integration of nanostructured metasurfaces for perturbed intensity sensing in SEIRAS has never been shown in combination with electrochemistry.

Here, we detect *in situ* the CO vibrational bending mode at 2033 cm$^{-1}$ emerging during the electrochemical conversion of CO into $CO_2$ using a platinum nano-slot metasurface on a $CaF_2$ substrate (**Figure 1a**) by coupling its resonance to the molecular vibrational mode and analysing the perturbation of the intensity in reflection. We investigated the linear CO vibrational mode ($CO_{linear}$) at 2033 cm$^{-1}$ because it is the most intense vibrational mode of CO on platinum[31–35]. The material of choice was platinum as it could fulfill all requirements, namely to function as a working electrode, support strong metasurface-driven resonances, and adsorb CO on its surface[36]. Moreover, Pt is a catalytic material for many reactions, making this platform very useful not only for the CO oxidation reaction but also for other reactions. The decision on the inverse structure (i.e. the slots), was made to preserve a connected metallic film that can carry electrical current. Moreover, compared to resonant rod-type antennas, the inverse counterparts have been shown to feature superior detection of molecular signal traces due to linearly instead of exponentially decaying EM near-fields[37]. The slots can only be excited with transverse electric (TE) polarized light[37]. We perform SEIRAS in an ATR geometry to further improve the sensing performance while maintaining free accessibility of the electrode surface for reactants and products[9,10]. We confirm the detection of adsorbed CO *via* the observation of the typical Stark shift and resolve a so far scarcely studied [38–40] effect due to the decrease of the CO coverage on the surface of platinum during the electrochemical oxidation. Furthermore, the presence of a second peak at 2086 cm$^{-1}$ on the spectral location of the linear vibrational mode could be attributed to the effect of the crystal orientation. Finally, we establish a methodology for designing similar nanophotonic-electrochemical platforms.

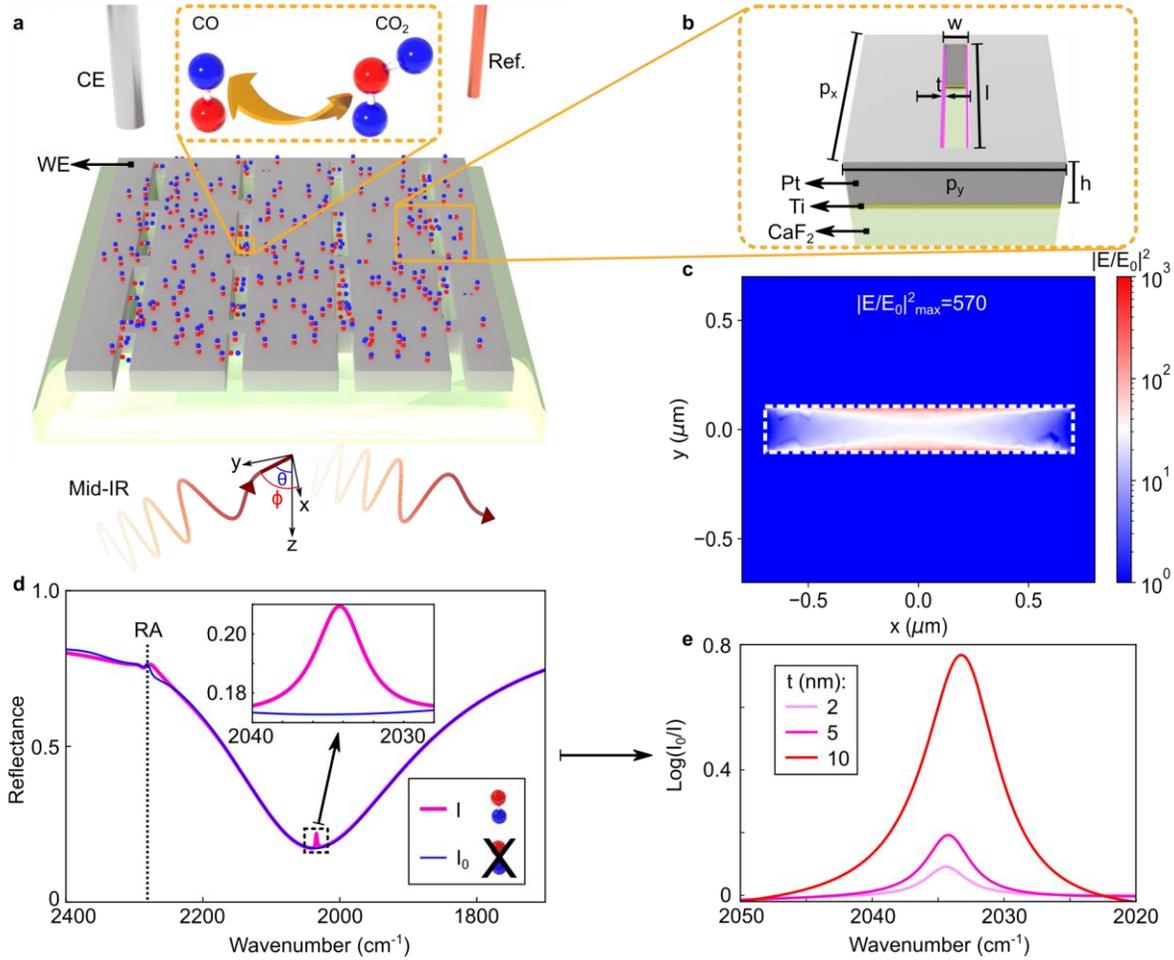

**Figure 1. Numerical design of the catalytic metasurface.** (a) Schematic for the Pt-based nano-slot metasurface for the *in-situ* integrated nanophotonic-electrochemical study of $CO_2$ oxidation. As the potential between the working electrode (WE) and the reference electrode (Ref.) is swept the presence of adsorbed CO is monitored via the detection of the linear vibrational mode of CO at 2033 cm$^{-1}$ with a Fourier transform infrared (FTIR) spectrometer. The nano-slot metasurface enhances the electromagnetic near-fields of TE polarized light in an ATR configuration coming in at an azimuthal angle $\phi = 0°$ and polar angle $\theta = 72°$ w.r.t. the Pt film ($xy$-plane). (b) Sketch of the Pt on $CaF_2$ nano-slot unit cell. Two CO model layers were included parallel to the long edges of the slot (magenta) with dimensions $l \times h \times t$. A 1 nm thick Ti adhesion layer was used in the fabrication of the structures but is not considered in the numerical simulations due to its negligible effect on the resonance position. The geometrical parameters of the unit cell for (c), (d), (e) are $h$ = 30 nm, $w$ = 200 nm, $l$ = 1380 nm, $p_y$ = 1400 nm, $p_x$ = 1600 nm. In (c) no CO model layer was included. In (d) $t$ = 5 nm. (c) Electric near field intensity (taken at $h$ = 30 nm) of the unit cell. The maximum near field intensity is 570. (d) The simulated reflectance spectrum of the metasurface with (pink) and without (blue) the CO model layer. The spectrum includes the Rayleigh anomaly (RA). (e) The differential absorbance of (d) with the thickness of the CO layer $t$ 2 nm (blue), 5 nm (pink) and 10 nm (red) showing clearly visible absorption bands.

**Results and discussion**

**Numerical design of catalytic nano-slot metasurface**

We start the implementation of our electrochemical sensing platform with the numerical design of the chosen nano-slot metasurface geometry. The structure consists of a unit cell composed of a single slot in an otherwise connected platinum film submerged in water on $CaF_2$ (**Figure 1b**). Notably, we model adsorbed CO by including an artificially created material covering the inside walls parallel to the long axis of the slot. The choice for the parameters of the unit cell was guided by Huck *et al.*[37] and modified in accordance with fabrication constraints. Huck *et al.*[37] optimized a gold nano-slot metasurface in the mid-infrared for normal incidence illumination in air for high quality factors (Q-factors) and electric near fields. The Q-factor relates the initial energy stored in a resonator to the energy dissipated in one radian of the cycle of oscillation[41].

On the basis of our simulations, the nano-slot metasurface achieves a resonance with a modulation in the absorbance of over 82% and a Q-factor of ca. 6.3 (see "Methods" section for details on the Q-factor calculation). Furthermore, the metasurface numerically exhibits an electric near-field enhancement $|E/E_0|^2$ of 570. This value can be increased in future experiments by decreasing the width of the slots[37] but was limited here due to fabrication constraints. The maximum electric near-field enhancement occurs inside the slots close to the faces parallel to its long axis (**Figure 1c**), with its electric field pointing orthogonally to it.

Huck *et al.*[37] found that the highest Q-factor and electric near field enhancement occurs when $w$ is small, $p_y = \lambda_{res}/2$, and $g = p_x - l = \lambda_{res}/2$, where $\lambda_{res}$ is the central wavelength of the resonance. However, to satisfy the experimental conditions the nano-slot metasurface was simulated in water instead of air and for an angle of incidence $\theta = 72°$. Under these conditions, tuning the resonance to 2033 cm$^{-1}$ ≈ 4.92 µm leads to the appearance of a Rayleigh anomaly (RA) such that $\lambda_{RA} > \lambda_{res}$, where $\lambda_{RA}$ is the central wavelength of the RA. The RA is a phenomenon associated with light diffracted parallel to the surface of a periodic structure[42]. When $\lambda_{RA} > \lambda_{res}$, the resonance lifetime and electric near-field enhancement is strongly reduced[43]. Consequently, a metasurface where $\lambda_{RA} > \lambda_{res}$ will exhibit poor sensing performance. For this reason, $g$ was reduced to 220 nm to push the resonance on the evanescent side of the RA (**Figure 1d**).

In coupled-resonator systems, the excitation efficiency of a resonator is significantly dependent on the ratio of its losses to external radiation $\gamma_e$, i.e. light scattering, and intrinsic material absorption $\gamma_i$ which strongly depends on the system design and parameters chosen[44]. When $\gamma_e \sim \gamma_i$ the system is critically-coupled and the second oscillator will lead to a dip in the absorption cross-section. SEIRAS performance can be maximized by utilizing a system that is close to the critical coupling condition[44,45]. Here, the nano-slot metasurface is near the critical-coupling condition with $\gamma_e/\gamma_i \approx$ 1.2. Thus, when the resonance overlaps with the vibrational mode of adsorbed CO at 2033 cm$^{-1}$ the coupling between the two resonators leads to a small peak in the reflectance spectrum (**Figure 1d**).

The idea that coverage effects and different analyte concentration can be sensed using our nano-slot metasurface is shown by changing the thickness of the model molecular layer representing adsorbed CO from 2, to 5, to 10 nm (**Figure 1e**). By increasing the thickness of the model molecular layer, a stronger differential absorbance $\log(I_0/I)$ can be obtained, where $I$ and $I_0$ are the reflectance measured with and without a CO model molecular layer, respectively (**Figure 1c**). Thus, a decrease in the coverage of an adsorbed material inside the slot can be linked to a decrease in the differential absorbance traces.

**Metasurface characterization**

First, the effect of the metasurface-driven resonance position on the coupling with $CO_{linear}$ vibration is studied in ATR mode using a focal plan array detector (**Figure 2a**). To test our nanophotonic-electrochemical platform, we first tuned the resonance position to match the $CO_{linear}$ vibration mode in 0.5M $K_2CO_3$ saturated with carbon monoxide. Then, we detuned the resonance to the blue and red spectral regions by decreasing and increasing the slot length $l$ by 200 nm from 1.33 µm, respectively (**Figure 2b**). There is a good fit between the numerically and experimentally obtained resonance positions, with a discrepancy of less than 40 cm$^{-1}$. As predicted by the simulations, a dip is observed in the resonance attributed to the $CO_{linear}$ vibration mode. Following these results, slots with a length of ca. 1.33 µm were found to match the $CO_{linear}$ vibrational mode on platinum. According to the literature[31–35,46] the $CO_{linear}$ mode should be spectrally located between 2020 cm$^{-1}$ and 2080 cm$^{-1}$. On the basis of our experiments, $CO_{linear}$ is located at 2033 cm$^{-1}$.

The differential absorbance highlights a more intense and well defined $CO_{linear}$ signal for the sample which has the best spectral overlap (**Figure 2c**). Two peaks can be observed at 2033 and 2086 cm$^{-1}$. The CO signals have a Fano-type line shape due to the narrow discrete $CO_{linear}$ vibration interfering with the broad spectral line of the metasurface-driven resonance[47]. The redshifted sample yields a highly asymmetric CO signal due to the strongly off-resonance coupling between the resonance and $CO_{linear}$[48]. In addition, the redshifted sample presents a strong peak around 1843 cm$^{-1}$, which is attributed to a second configuration of adsorption, the CO bridge ($CO_{bridge}$)[31,32,34,46]. The scanning electron microscopy images show good quality of the fabricated nanostructures (**Figure 2d**). For the next part of this work, the electrochemical behavior of the sample with matching spectral overlap of its resonance with the $CO_{linear}$ vibrational mode is studied.

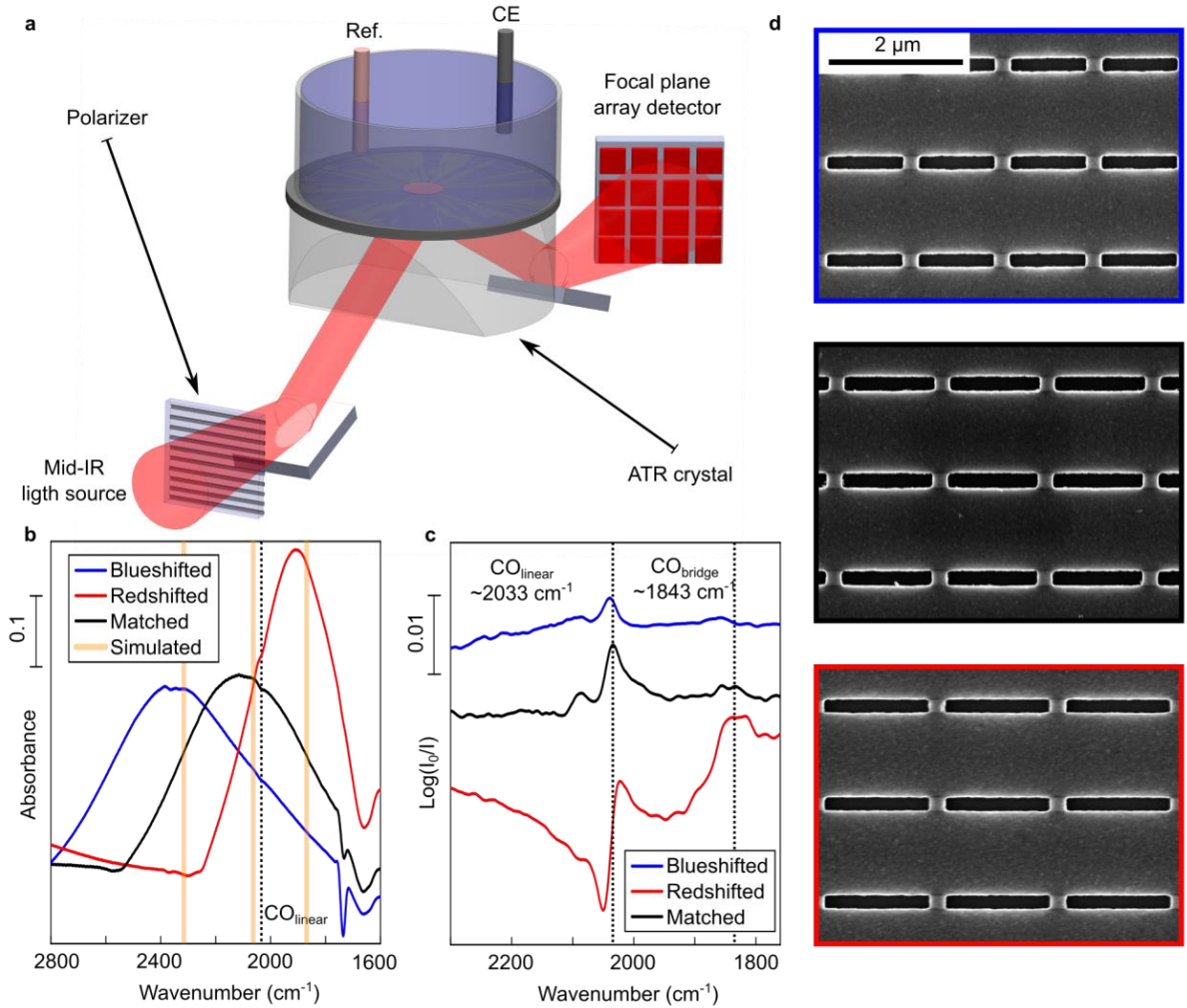

**Figure 2. Testing the nanophotonic-electrochemical platform.** (a) A schematic showing the experimental setup used to perform SEIRAS in an ATR geometry. A continuous mid-IR collimated linearly polarized light source illuminated the nano-slot metasurface from below at an angle of ca. 72º. A focal plane detector array was used to collect the signal. (b) Experimental absorbance spectra of three Pt nano-slot metasurface in CO saturated 0.5M $K_2CO_3$ aqueous electrolyte interfaces with resonance positions around the ideal position for $CO_{linear}$ with slot lengths $l$ 1.33 µm (black, ideal) and two detuned resonance positions with slots length of 1.13 µm (blue) and 1.53 µm (red). The other parameters are $h$ = 30 nm, $w$ = 200 nm, $p_y$ = 1440 nm, gap = $p_x - l$ = 220 nm. The resonance position of $CO_{linear}$ vibration is indicated by the black dahsed line. The numerically modelled resonance positions at 2312 cm$^{-1}$, 2066 cm$^{-1}$, and 1876 cm$^{-1}$ (yellow) are shown for comparison. (c) The differential absorbance of the CO signal after baseline correction for the blueshifted, matched, and redshifted resonances. (d) Scanning electron microscopy images corresponding to the nano-slot metasurfaces used in (b) and (c).

**CO adsorption at Open Circuit Potential**

Here, we follow *in situ* the CO adsorption during the saturation of an electrolyte at open circuit potential (OCP) and characterize the CO adsorption by performing SEIRAS concurrently with electrochemical cyclic voltammetry. The transition from the Ar to the CO saturated electrolyte is accompanied by a shift of the OCP due to a change of the equilibrium determining redox reaction (**Figure 3a**). At the equilibrium potential of the Ar saturated electrolyte (ca. 300 mV$_{SHE}$) CO is oxidized and the OCP drops towards negative values where CO adsorbs on the Pt surface.

The SEIRAS measurements were taken in 0.5M K$_2$CO$_3$ with and without CO (**Figure 3b**). Looking at the differential absorbance (**Figure 3c**), a distortion of the base line appears at 2460 seconds (+25 mV$_{SHE}$). Then, after ca. 2800 seconds (-270 mV$_{SHE}$) two clearly distinguishable CO$_{linear}$ peaks emerge. These peaks become more discernible with time as the coverage of adsorbed CO increases. As the intensity of the peaks stabilizes the maximum coverage of CO is reached.

The CO signal obtained with the nano-slot metasurface compared to that obtained with a pure platinum layer (30 nm) at the OCP is increased by an estimated factor of 27 (**Figure 3d**). Both samples have been evaporated simultaneously. This gives both systems the same material properties such as the surface roughness. For this reason, the 27-fold difference between the signals obtained with the two systems can be directly linked to the metasurfaces-driven enhancement provided by nanostructuring the surface of the working electrode.

For adsorbed CO to interact with incident light, the orientation of the transition dipole moment of the CO vibrational mode relative to the electric field component needs to be non-zero[49]. Consequently, only the (interior) side walls parallel to the long axis of the slots can be considered active representing a ratio of active to total surface of 3.6% compared to a smooth platinum layer. This leads to an experimentally determined local signal enhancement of above 700.

The second peak at 2086 cm$^{-1}$ was only observed using the nano-slot metasurface. The most likely explanation could be that the higher resolution achieved with the nano-slot metasurface allows for the deconvolution of this peak from the background, which was not possible in previous architectures based on a continuous Pt film. According to the literature, several possibilities exist. The first assumption is that CO could adsorbs on different crystal orientations with different binding energies[46,50]. As reported by A. Cuesta *et al.*[50], an adsorption on Pt(111) single crystals was found at around 2070 cm$^{-1}$ [38,50,51], while CO adsorbed on Pt(100) electrodes was detected between 2027 cm$^{-1}$ [52,53] and 2050 cm$^{-1}$ [50,54]. These two values are in good agreement with the ones observed here (2086 cm$^{-1}$ and 2033 cm$^{-1}$). Another possibility is the adsorption of CO on terraces (higher frequency band at 2086 cm$^{-1}$), steps and defects (lower frequency band at 2033 cm$^{-1}$) [31,32,55].

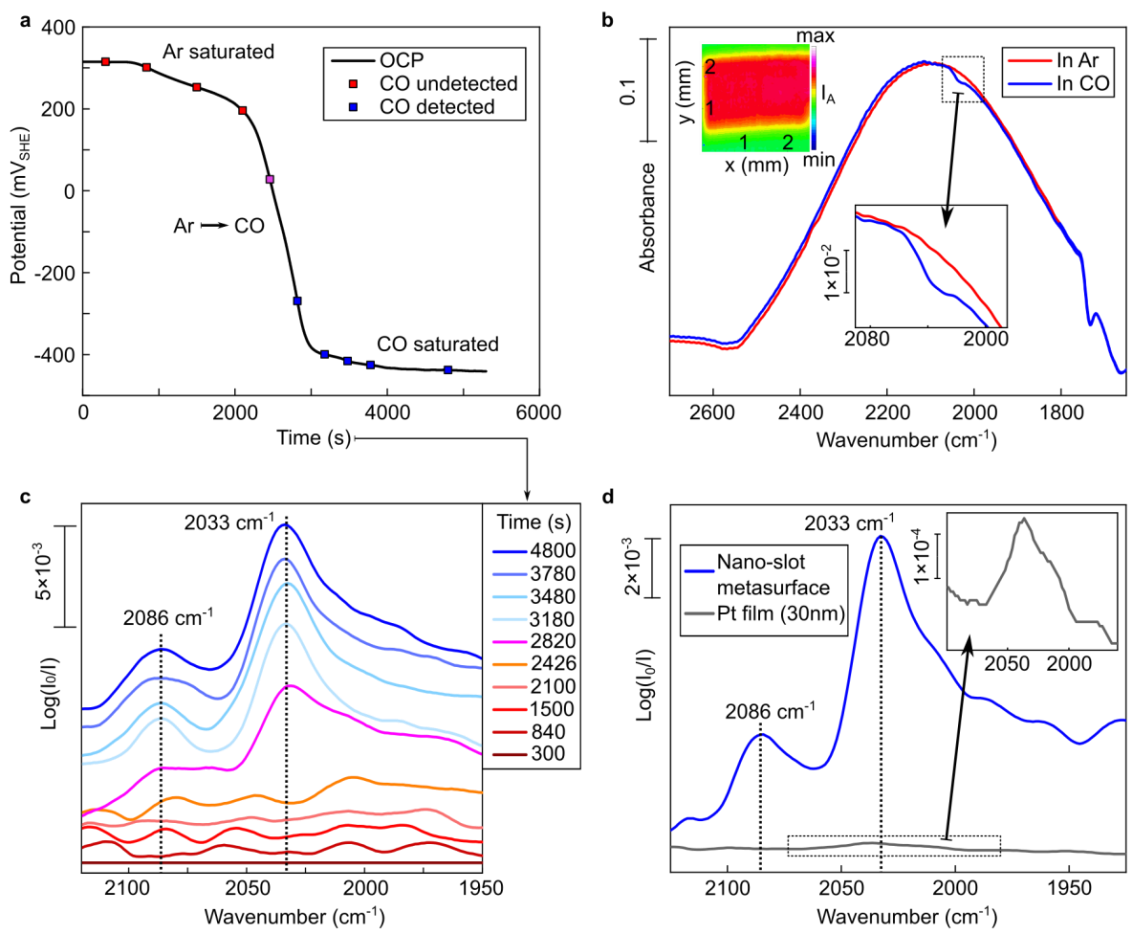

**Figure 3. Electrochemical and spectroscopic response of the nanophotonic platform at the OCP during a gas transition of from an Ar-saturated electrolyte to a CO-saturated one.** (a) The evolution of the OCP of the platinum nano-slot metasurface during the transition from an Ar saturated ($Ar^{sat}$ around +310 $mV_{SHE}$) to a CO saturated electrolyte ($CO^{sat}$ around -440 $mV_{SHE}$). (b) FTIR spectra of the Pt nano-slot metasurface/electrolyte interface in $Ar^{sat}$ and $CO^{sat}$. The heat map represents the integrated area below the resonance between 2600 and 1800 $cm^{-1}$ collected by an array of 64 by 64 detectors. (c) The evolution of the differential absorbance $CO_{linear}$ peaks during the CO bubbling process. (d) Comparison of $CO_{linear}$ signals obtained in $CO^{sat}$ after 80 min of CO bubbling with a pure Pt layer and with the nanophotonic-electrochemical platform.

**CO oxidation on platinum**

The behavior of the nano-slot metasurface was evaluated during the electrochemical oxidation of carbon monoxide using electrochemical cyclic voltammetry. The anodic scan in CO-saturated electrolyte (CO$^{sat}$) presents an initial state with a low current (**Figure 4a**, black line). When an applied potential of around -150 mV$_{SHE}$ is reached, the current density plateaus at around +25 µA.cm$^{-2}$, which is attributed to CO oxidation[56,57]. At ca. 450 mV$_{SHE}$ the current density starts to decrease. The origin of this decrease is still debated in the literature. One explanation attributes the decreasing current density to competing adsorption of CO and OH on the Pt surface at higher potentials[58]. Another possibility discussed is that the formation of a thin oxide or hydroxide Pt layer prevents the oxidation of CO. The latter assumption is supported by the reduction dip (from +160 to -80 mV$_{SHE}$) of platinum in Argon saturated electrolyte (Ar$^{sat}$) (**Figure 4a-b**). The behavior of the cathodic scan is similar, except that the onset of CO oxidation is shifted to more negative potentials resulting in a hysteresis. Moreover, a shift in the onset of the hydrogen evolution reaction (HER) in Ar$^{sat}$ and CO$^{sat}$ electrolytes is observed, highlighting the poisoning behavior of adsorbed CO on the platinum surface[59].

Similarly to our Fourier-transform infrared (FTIR) measurements in CO$^{sat}$ electrolyte under OCP (**Figure 3c**), two CO$_{linear}$ peaks were also found during the electrochemical potential sweeps (**Figure 4c-d**). There is a spectral shift during the anodic and cathodic scan (between -650 and -150 mV$_{SHE}$) which is attributed to either a higher π-back-donation from the metal to CO[40,60] and/or to the Stark effect. The Stark effect results from the interactions between the surface electric field and the dipole moment of the adsorbates[60–62]. During the anodic scan (**Figure 4e**), the most intense peak shows a blue-shift of 53 cm$^{-1}$/V in agreement with the literature[38–40,61,63]. The second peak shows a blueshift of 33 cm$^{-1}$/V. Between -50 and +50 mV$_{SHE}$ a redshift is observed which is not well documented in the literature[39,55,60,64]. The redshift is attributed to a decrease of the CO coverage due to its oxidation into $CO_2$, decreasing the dipole-dipole interactions[55,65]. The observation of the coverage effect was possible here due to the high resolution reached with the nano-slot metasurface. It was not resolved with a continuous platinum film. During the anodic scan, there is a slight increase in the area of the first peak (~2033 cm$^{-1}$), while the area of the second peak slightly decreases. This behavior could be explained by a surface migration of adsorbed CO to a more stable position[33,39,40,52]. Alternatively, the reconstruction or roughening of the Pt surface with electrical polarization[66,67] could lead to a modification of the surface microstructure and CO adsorption energy[68]. Looking at spectra obtained during the cathodic scan (**Figure 4f**), the second peak (~2086 cm$^{-1}$) almost disappeared. This supports the assumption that the cause is the platinum surface modification at high applied potentials. At high cathodic potentials (-550 to -650 mV$_{SHE}$) a decrease of the CO peak is observed and attributed to the HER[63], indicating that the adsorption of hydrogen displaces adsorbed CO.

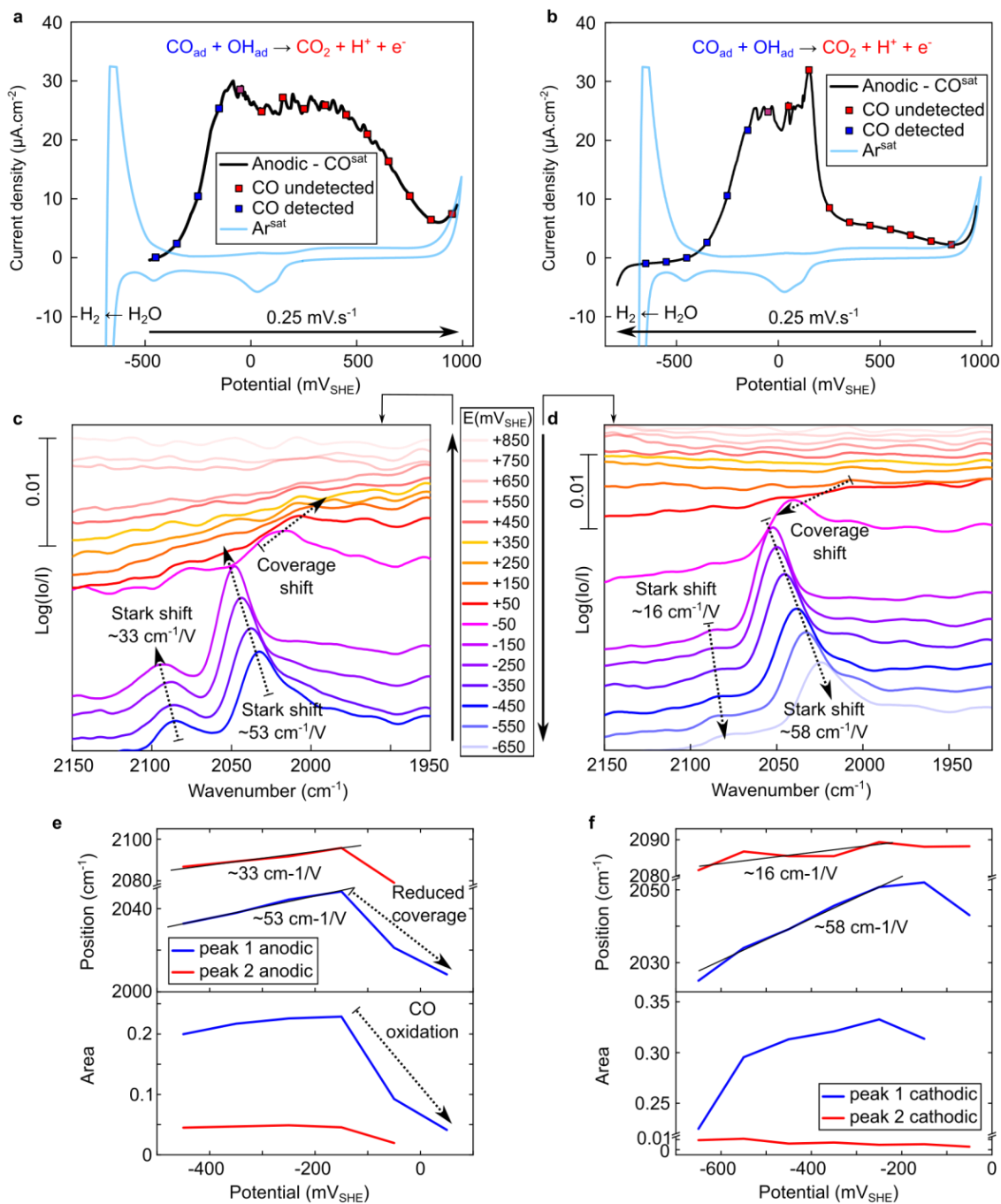

**Figure 4. Behavior of the Pt nano-slot metasurface during cyclic voltammetry in 0.5M $K_2CO_3$ saturated with CO at 0.25 mV.s$^{-1}$.** Evolution of the current density with the potential during the (a) anodic (from OCP to +1000 mV$_{SHE}$) and (b) cathodic (from +1000 mV$_{SHE}$ to -700 mV$_{SHE}$) scan in CO$^{sat}$ electrolyte (black line). For comparison, the blue line depicts CVs in an Ar$^{sat}$ electrolyte. Evolution of SEIRAS spectra with, the (c) anodic and (d) cathodic scans acquired every 100 mV. Evolution of the position and area of the CO$_{linear}$ peak during the (e) anodic and (f) cathodic scans.

**Conclusion**

To the best of our knowledge, we have developed the first hybrid nanophotonic-electrochemical platform for SEIRAS based on a platinum nano-slot metasurface. The resonance of the metasurface was numerically modelled giving a maximum electric near-field intensity enhancement of 570. The resonance was tuned to couple with and enhance the CO vibrational mode at 2033 cm$^{-1}$. The principle behind the sensing improvement due to the electric near-field enhancement was tested by fabricating on-resonance and detuned metasurfaces and carefully analyzing the resonance. The numerical simulations and SEIRAS experimental results were in good agreement. Two peaks were resolved for the CO$_{linear}$ mode which could be attributed to adsorption of CO on Pt(111) and Pt(100). CO$_{linear}$ was best observed with a spectrally overlapping resonance leading to an experimental signal improvement of more than 27 over a conventionally used platinum film. During the electrochemical oxidation of CO, a classic Stark effect was observed. Moreover, thanks to the high resolution provided by the nano-slot metasurface, a redshift of CO$_{linear}$ was observed, linked to a decrease of the coverage of adsorbed CO due to its oxidation. We anticipate our proof-of-concept nanophotonic-electrochemical platform for SEIRAS to guide new system designs and material combinations suitable to characterize different electrochemical interfaces, reaction products, and short-lived intermediates.

**Methods**

**Numerical simulations**

The simulations were performed in CST Studio Suite 2021 using the finite-element frequency-domain Maxwell solver. CaF$_2$ was simulated using a refractive index, n, of 1.4, the surrounding medium as water with n ≈ 1.33 and platinum using the data given by Rakić *et al.*[69] The inside walls perpendicular to the electric field were covered with a model material to represent the adsorbed $CO_{linear}$ vibrational mode at ~2033 cm$^{-1}$ (see Supplementary Information for more details). The titanium adhesion layer was not simulated as including it did not lead to substantial spectral shifts in the resonance position. An impedance-matched open port with a perfectly matched layer introduced linearly polarized light at an angle of 72° across the CaF$_2$ layer towards the nano-slots. At 72° the light was totally internally reflected at the CaF$_2$-Pt interface. Therefore, the boundary opposite the open port was set as perfect electric conductor. The unit cell was defined and then simulated as infinite periodic array *via* Floquet boundaries. A field monitor was placed at the center of the slot in the xy-plane. The highest field enhancement is found slightly above the apex of the slots. The value of the highest field enhancement of the system was evaluated within the volume of the numerical model. To extract the Q-factor and coupling ratio $\gamma_e/\gamma_i$, the simulated resonance was fitted in reflectance (**Figure 1d**, blue curve) using temporal coupled mode theory according to Hu *et al.*[70]

**Metasurface fabrication**

CaF$_2$ was selected as the substrate due to its transparent nature in the mid-IR spectral range, low solubility, and high chemical stability. The measurements shown in **Figures 3** and **4** used metasurface arrays were with at least 2200 by 2700 unit cells resulting in a pattern area of approximately 13.3 mm$^2$, which ensured that there are more than enough unit cells for the measured resonance to correspond to the mode of the infinite periodic array used for the numerical simulations. After sample cleaning (acetone bath in an ultrasonic cleaner followed by oxygen plasma cleaning) the substrate was spin-coated first with an adhesion promoter (Surpass 4000), then with a layer of negative tone photoresist (ma-N 2403) which was baked at 100°C for 60s, and finally with a conducting layer (ESpacer 300Z). The metasurface patterns were written *via* electron-beam lithography (Raith Eline Plus) with an acceleration voltage of 30 kV and an aperture of 20 µm. The exposed resist was developed in ma-D 525 for 70s at room temperature. The patterned surface was then coated with a titanium adhesion layer (1 nm at 0.4 Å/s) and a platinum film (30 nm at 2 Å/s) using electron-beam evaporation (PRO Line PVD 75, Lesker). Finally, an overnight lift-off in mr-REM 700 concluded the top-down fabrication process. A pure 30 nm thick platinum film on 1 nm titanium on $CaF_2$ functioned as reference for the *in-situ* SEIRAS measurements.

### *In-situ* SEIRAS and electrochemical measurements

SEIRAS was performed using a Vertex 80 coupled with an IMAC chamber from Bruker. Each sample was mounted on a VeeMax III (purged with $N_2$) from PIKE Technologies in attenuated total internal reflection (ATR) mode with a light polarizer, an electrochemical Jackfish cell and a $CaF_2$ prism bevelled at 72°. A classical three electrode system was used with a Saturated Calomel Electrode (E= 0.244 $V_{SHE}$), a platinum wire as counter electrode and the platinum sample as working electrode. The IMAC chamber is equipped with a focal plane array detector composed of 64 x 64 MCT-detectors (a total of 4096 detectors), which allows to perform a mapping of the studied sample. Each detector collects its own spectrum and then, the active slot covered area is detected by integration of each spectrum between 1600 to 2800 cm$^{-1}$ (**Figure 3b**, inset). Finally, an average can be determined using the spectra of the detectors that probed the resonance. A baseline correction is applied to this average as well as a Savitzky–Golay filter to smoothen the data.

For the characterization of the resonance, its position was determined using three samples composed of arrays with different nanostructure sizes. For each sample, the resonance was measured in 0.5M $K_2CO_3$ electrolyte saturated with Ar and then saturated with CO. Prior to the first characterization, a cyclic voltammogram (20 mV.s$^{-1}$) was recorded in order to confirm the cleanliness of the electrode surface. Then, an initial background was acquired using p-polarized light and the Fano resonance was characterized using s-polarized light. Each spectrum was recorded with a resolution of 4 cm$^{-1}$ and the final mapping results from a collection of 32 scans. The enhancement of the nano-slot metasurface is obtained by comparison of the CO$_{linear}$ vibrational mode on a pure Pt layer (30nm) without nanostructures.

The adsorption of CO during the transition from Ar$^{sat}$ to CO$^{sat}$ electrolyte (0.5M K$_2$CO$_3$) was studied using the nano-slot metasurface with the best overlapping resonance with the CO$_{linear}$ vibration mode. Cleaning and background acquisition protocol were the same as described above. Carbon monoxide was slowly flowed into the electrochemical cell and spectra were acquired regularly during the transition from Ar$^{sat}$ to CO$^{sat}$ at the OCP.

The oxidation of CO during potential sweeps was investigated after 2 hours of CO bubbling. A cyclic voltammogram, with a slow scan rate (0.25 mV.s$^{-1}$), from the OCP to +1000 mV$_{SHE}$ and back to -760 mV$_{SHE}$ was performed and a spectrum was acquired every 100 mV.

**References**

(1) Lagadec, M. F.; Grimaud, A. Water Electrolysers with Closed and Open Electrochemical Systems. *Nat. Mater.* **2020**, *19* (11), 1140–1150. https://doi.org/10.1038/s41563-020-0788-3.
(2) Sullivan, I.; Goryachev, A.; Digdaya, I. A.; Li, X.; Atwater, H. A.; Vermaas, D. A.; Xiang, C. Coupling Electrochemical CO2 Conversion with CO2 Capture. *Nat. Catal.* **2021**, *4* (11), 952–958. https://doi.org/10.1038/s41929-021-00699-7.
(3) Stephens, I. E. L.; Chan, K.; Bagger, A.; Boettcher, S. W.; Bonin, J.; Boutin, E.; Buckley, A. K.; Buonsanti, R.; Cave, E. R.; Chang, X.; Chee, S. W.; Silva, A. H. M. da; Luna, P. de; Einsle, O.; Endrődi, B.; Escudero-Escribano, M.; Araujo, J. V. F. de; Figueiredo, M. C.; Hahn, C.; Hansen, K. U.; Haussener, S.; Hunegnaw, S.; Huo, Z.; Hwang, Y. J.; Janáky, C.; Jayathilake, B. S.; Jiao, F.; Jovanov, Z. P.; Karimi, P.; Koper, M. T. M.; Kuhl, K. P.; Lee, W. H.; Liang, Z.; Liu, X.; Ma, S.; Ma, M.; Oh, H.-S.; Robert, M.; Cuenya, B. R.; Rossmeisl, J.; Roy, C.; Ryan, M. P.; Sargent, E. H.; Sebastián-Pascual, P.; Seger, B.; Steier, L.; Strasser, P.; Varela,

A. S.; Vos, R. E.; Wang, X.; Xu, B.; Yadegari, H.; Zhou, Y. 2022 Roadmap on Low Temperature Electrochemical CO2 Reduction. *J. Phys. Energy* **2022**, *4* (4), 042003. https://doi.org/10.1088/2515-7655/ac7823.

(4) Zhao, S.; Jin, R.; Jin, R. Opportunities and Challenges in $CO_2$ Reduction by Gold- and Silver-Based Electrocatalysts: From Bulk Metals to Nanoparticles and Atomically Precise Nanoclusters. *ACS Energy Lett.* **2018**, *3* (2), 452–462. https://doi.org/10.1021/acsenergylett.7b01104.

(5) Wuttig, A.; Yaguchi, M.; Motobayashi, K.; Osawa, M.; Surendranath, Y. Inhibited Proton Transfer Enhances Au-Catalyzed $CO_2$-to-Fuels Selectivity. *Proc. Natl. Acad. Sci.* **2016**, *113* (32), E4585–E4593. https://doi.org/10.1073/pnas.1602984113.

(6) Hammes, G. G. *Spectroscopy for the Biological Sciences*; John Wiley & Sons, 2005.

(7) Bernath, P. F. *Spectra of Atoms and Molecules*; Oxford University Press, USA, 2005.

(8) Lotti, D.; Hamm, P.; Kraack, J. P. Surface-Sensitive Spectro-Electrochemistry Using Ultrafast 2D ATR IR Spectroscopy. *J. Phys. Chem. C* **2016**, *120* (5), 2883–2892. https://doi.org/10.1021/acs.jpcc.6b00395.

(9) Yang, X.; Sun, Z.; Low, T.; Hu, H.; Guo, X.; García de Abajo, F. J.; Avouris, P.; Dai, Q. Nanomaterial-Based Plasmon-Enhanced Infrared Spectroscopy. *Adv. Mater.* **2018**, *30* (20), 1704896. https://doi.org/10.1002/adma.201704896.

(10) Neubrech, F.; Huck, C.; Weber, K.; Pucci, A.; Giessen, H. Surface-Enhanced Infrared Spectroscopy Using Resonant Nanoantennas. *Chem. Rev.* **2017**, *117* (7), 5110–5145. https://doi.org/10.1021/acs.chemrev.6b00743.

(11) Ribeiro, J. A.; Sales, M. G. F.; Pereira, C. M. Electrochemistry Combined-Surface Plasmon Resonance Biosensors: A Review. *TrAC Trends Anal. Chem.* **2022**, *157*, 116766. https://doi.org/10.1016/j.trac.2022.116766.

(12) Vijay, S.; Hogg, T. V.; Ehlers, J.; Kristoffersen, H. H.; Katayama, Y.; Shao Horn, Y.; Chorkendorff, I.; Chan, K.; Seger, B. Interaction of CO with Gold in an Electrochemical Environment. *J. Phys. Chem. C* **2021**, *125* (32), 17684–17689. https://doi.org/10.1021/acs.jpcc.1c04013.

(13) Katayama, Y.; Nattino, F.; Giordano, L.; Hwang, J.; Rao, R. R.; Andreussi, O.; Marzari, N.; Shao-Horn, Y. An In Situ Surface-Enhanced Infrared Absorption Spectroscopy Study of Electrochemical CO2 Reduction: Selectivity Dependence on Surface C-Bound and O-Bound Reaction Intermediates. *J. Phys. Chem. C* **2019**, *123* (10), 5951–5963. https://doi.org/10.1021/acs.jpcc.8b09598.

(14) Dunwell, M.; Lu, Q.; Heyes, J. M.; Rosen, J.; Chen, J. G.; Yan, Y.; Jiao, F.; Xu, B. The Central Role of Bicarbonate in the Electrochemical Reduction of Carbon Dioxide on Gold. *J. Am. Chem. Soc.* **2017**, *139* (10), 3774–3783. https://doi.org/10.1021/jacs.6b13287.

(15) Wang, D.; Shi, F.; Jose, J.; Hu, Y.; Zhang, C.; Zhu, A.; Grzeschik, R.; Schlücker, S.; Xie, W. In Situ Monitoring of Palladium-Catalyzed Chemical Reactions by Nanogap-Enhanced Raman Scattering Using Single Pd Cube Dimers. *J. Am. Chem. Soc.* **2022**, *144* (11), 5003–5009. https://doi.org/10.1021/jacs.1c13240.

(16) Rodrigo, D.; Tittl, A.; Ait-Bouziad, N.; John-Herpin, A.; Limaj, O.; Kelly, C.; Yoo, D.; Wittenberg, N. J.; Oh, S.-H.; Lashuel, H. A.; Altug, H. Resolving Molecule-Specific Information in Dynamic Lipid Membrane Processes with Multi-Resonant Infrared Metasurfaces. *Nat. Commun.* **2018**, *9* (1), 2160. https://doi.org/10.1038/s41467-018-04594-x.

(17) Tittl, A.; Leitis, A.; Liu, M.; Yesilkoy, F.; Choi, D.-Y.; Neshev, D. N.; Kivshar, Y. S.; Altug, H. Imaging-Based Molecular Barcoding with Pixelated Dielectric Metasurfaces. *Science* **2018**, *360* (6393), 1105–1109. https://doi.org/10.1126/science.aas9768.
(18) Hosseini Farahabadi, S. A.; Entezami, M.; Abouali, H.; Amarloo, H.; Poudineh, M.; Safavi-Naeini, S. Sub-Terahertz Silicon-Based on-Chip Absorption Spectroscopy Using Thin-Film Model for Biological Applications. *Sci. Rep.* **2022**, *12* (1), 17747. https://doi.org/10.1038/s41598-022-21015-8.
(19) Li, T.; Shang, D.; Gao, S.; Wang, B.; Kong, H.; Yang, G.; Shu, W.; Xu, P.; Wei, G. Two-Dimensional Material-Based Electrochemical Sensors/Biosensors for Food Safety and Biomolecular Detection. *Biosensors* **2022**, *12* (5), 314. https://doi.org/10.3390/bios12050314.
(20) Wang, J.; Maier, S. A.; Tittl, A. Trends in Nanophotonics-Enabled Optofluidic Biosensors. *Adv. Opt. Mater.* **2022**, *10* (7), 2102366. https://doi.org/10.1002/adom.202102366.
(21) Jiang, J.; Wang, X.; Li, S.; Ding, F.; Li, N.; Meng, S.; Li, R.; Qi, J.; Liu, Q.; Liu, G. L. Plasmonic Nano-Arrays for Ultrasensitive Bio-Sensing. *Nanophotonics* **2018**, *7* (9), 1517–1531. https://doi.org/10.1515/nanoph-2018-0023.
(22) Li, N.; Lu, Y.; Li, S.; Zhang, Q.; Wu, J.; Jiang, J.; Liu, G. L.; Liu, Q. Monitoring the Electrochemical Responses of Neurotransmitters through Localized Surface Plasmon Resonance Using Nanohole Array. *Biosens. Bioelectron.* **2017**, *93*, 241–249. https://doi.org/10.1016/j.bios.2016.08.105.
(23) Nakamoto, K.; Kurita, R.; Niwa, O. Electrochemical Surface Plasmon Resonance Measurement Based on Gold Nanohole Array Fabricated by Nanoimprinting Technique. *Anal. Chem.* **2012**, *84* (7), 3187–3191. https://doi.org/10.1021/ac203160r.
(24) Wang, X.; Huang, S.-C.; Hu, S.; Yan, S.; Ren, B. Fundamental Understanding and Applications of Plasmon-Enhanced Raman Spectroscopy. *Nat. Rev. Phys.* **2020**, *2* (5), 253–271. https://doi.org/10.1038/s42254-020-0171-y.
(25) Wright, D.; Lin, Q.; Berta, D.; Földes, T.; Wagner, A.; Griffiths, J.; Readman, C.; Rosta, E.; Reisner, E.; Baumberg, J. J. Mechanistic Study of an Immobilized Molecular Electrocatalyst by in Situ Gap-Plasmon-Assisted Spectro-Electrochemistry. *Nat. Catal.* **2021**, *4* (2), 157–163. https://doi.org/10.1038/s41929-020-00566-x.
(26) Fang, Y.; Wang, H.; Yu, H.; Liu, X.; Wang, W.; Chen, H.-Y.; Tao, N. J. Plasmonic Imaging of Electrochemical Reactions of Single Nanoparticles. *Acc. Chem. Res.* **2016**, *49* (11), 2614–2624. https://doi.org/10.1021/acs.accounts.6b00348.
(27) Cheng, X. R.; Wallace, G. Q.; Lagugné-Labarthet, F.; Kerman, K. Au Nanostructured Surfaces for Electrochemical and Localized Surface Plasmon Resonance-Based Monitoring of α-Synuclein–Small Molecule Interactions. *ACS Appl. Mater. Interfaces* **2015**, *7* (7), 4081–4088. https://doi.org/10.1021/am507972b.
(28) Wang, S.-S.; Zhao, X.-P.; Liu, F.-F.; Younis, M. R.; Xia, X.-H.; Wang, C. Direct Plasmon-Enhanced Electrochemistry for Enabling Ultrasensitive and Label-Free Detection of Circulating Tumor Cells in Blood. *Anal. Chem.* **2019**, *91* (7), 4413–4420. https://doi.org/10.1021/acs.analchem.8b04908.
(29) Cortés, E.; Wendisch, F. J.; Sortino, L.; Mancini, A.; Ezendam, S.; Saris, S.; Menezes, L. de S.; Tittl, A.; Ren, H.; Maier, S. A. Optical Metasurfaces for Energy Conversion. *Chem. Rev.* **2022**, *122* (19), 15082–15176. https://doi.org/10.1021/acs.chemrev.2c00078.
(30) Stefancu, A.; Nan, L.; Zhu, L.; Chiș, V.; Bald, I.; Liu, M.; Leopold, N.; Maier, S. A.; Cortes, E. Controlling Plasmonic Chemistry Pathways through Specific Ion Effects. *Adv. Opt. Mater.* **2022**, *10* (14), 2200397. https://doi.org/10.1002/adom.202200397.

(31) Hossain, M. J.; Rahman, M. M.; Jafar Sharif, Md. Preference for Low-Coordination Sites by Adsorbed CO on Small Platinum Nanoparticles. *Nanoscale Adv.* **2020**, *2* (3), 1245–1252. https://doi.org/10.1039/C9NA00499H.

(32) Silva, C. D.; Cabello, G.; Christinelli, W. A.; Pereira, E. C.; Cuesta, A. Simultaneous Time-Resolved ATR-SEIRAS and CO-Charge Displacement Experiments: The Dynamics of CO Adsorption on Polycrystalline Pt. *J. Electroanal. Chem.* **2017**, *800*, 25–31. https://doi.org/10.1016/j.jelechem.2016.10.034.

(33) McPherson, I. J.; Ash, P. A.; Jones, L.; Varambhia, A.; Jacobs, R. M. J.; Vincent, K. A. Electrochemical CO Oxidation at Platinum on Carbon Studied through Analysis of Anomalous in Situ IR Spectra. *J. Phys. Chem. C* **2017**, *121* (32), 17176–17187. https://doi.org/10.1021/acs.jpcc.7b02166.

(34) Susarrey-Arce, A.; Tiggelaar, R. M.; Gardeniers, J. G. E.; van Houselt, A.; Lefferts, L. CO Adsorption on Pt Nanoparticles in Low E-Fields Studied by ATR-IR Spectroscopy in a Microreactor. *J. Phys. Chem. C* **2015**, *119* (44), 24887–24894. https://doi.org/10.1021/acs.jpcc.5b08392.

(35) Smolinka, T.; Heinen, M.; Chen, Y. X.; Jusys, Z.; Lehnert, W.; Behm, R. J. CO2 Reduction on Pt Electrocatalysts and Its Impact on H2 Oxidation in CO2 Containing Fuel Cell Feed Gas – A Combined in Situ Infrared Spectroscopy, Mass Spectrometry and Fuel Cell Performance Study. *Electrochimica Acta* **2005**, *50* (25–26), 5189–5199. https://doi.org/10.1016/j.electacta.2005.02.082.

(36) Katayama, Y.; Giordano, L.; Rao, R. R.; Hwang, J.; Muroyama, H.; Matsui, T.; Eguchi, K.; Shao-Horn, Y. Surface (Electro)Chemistry of CO2 on Pt Surface: An in Situ Surface-Enhanced Infrared Absorption Spectroscopy Study. *J. Phys. Chem. C* **2018**, *122* (23), 12341–12349. https://doi.org/10.1021/acs.jpcc.8b03556.

(37) Huck, C.; Vogt, J.; Sendner, M.; Hengstler, D.; Neubrech, F.; Pucci, A. Plasmonic Enhancement of Infrared Vibrational Signals: Nanoslits versus Nanorods. *ACS Photonics* **2015**, *2* (10), 1489–1497. https://doi.org/10.1021/acsphotonics.5b00390.

(38) Villegas, I.; Weaver, M. J. Carbon Monoxide Adlayer Structures on Platinum (111) Electrodes: A Synergy between *In-situ* Scanning Tunneling Microscopy and Infrared Spectroscopy. *J. Chem. Phys.* **1994**, *101* (2), 1648–1660. https://doi.org/10.1063/1.467786.

(39) Watanabe, S.; Inukai, J.; Ito, M. Coverage and Potential Dependent CO Adsorption on Pt(111), (711) and (100) Electrode Surfaces Studied by Infrared Reflection Absorption Spectroscopy. **2020**, 9.

(40) Nakamura, M.; Ogasawara, H.; Inukai, J.; Ito, M. CO Migration on Pt(100) and Pt(11 1 1) Surfaces Studied by Time Resolved Infrared Reflection-Absorption Spectroscopy. *Surf. Sci.* **1993**, *283* (1–3), 248–254. https://doi.org/10.1016/0039-6028(93)90989-W.

(41) Hickman, I. *Analog Electronics: Analog Circuitry Explained*; Newnes, 2013.

(42) Gao, H.; McMahon, J. M.; Lee, M. H.; Henzie, J.; Gray, S. K.; Schatz, G. C.; Odom, T. W. Rayleigh Anomaly-Surface Plasmon Polariton Resonances in Palladium and Gold Subwavelength Hole Arrays. *Opt. Express* **2009**, *17* (4), 2334. https://doi.org/10.1364/OE.17.002334.

(43) Adato, R.; Yanik, A. A.; Wu, C.-H.; Shvets, G.; Altug, H. Radiative Engineering of Plasmon Lifetimes in Embedded Nanoantenna Arrays. *Opt. Express* **2010**, *18* (5), 4526. https://doi.org/10.1364/OE.18.004526.

(44) Adato, R.; Artar, A.; Erramilli, S.; Altug, H. Engineered Absorption Enhancement and Induced Transparency in Coupled Molecular and Plasmonic Resonator Systems. *Nano Lett.* **2013**, *13* (6), 2584–2591. https://doi.org/10.1021/nl400689q.

(45) Aigner, A.; Tittl, A.; Wang, J.; Weber, T.; Kivshar, Y.; Maier, S. A.; Ren, H. Plasmonic Bound States in the Continuum to Tailor Light-Matter Coupling. *Sci. Adv.* **2022**, *8* (49), eadd4816. https://doi.org/10.1126/sciadv.add4816.

(46) Farias, M. J. S.; Busó-Rogero, C.; Tanaka, A. A.; Herrero, E.; Feliu, J. M. Monitoring of CO Binding Sites on Stepped Pt Single Crystal Electrodes in Alkaline Solutions by in Situ FTIR Spectroscopy. *Langmuir* **2020**, *36* (3), 704–714. https://doi.org/10.1021/acs.langmuir.9b02928.

(47) Luk'yanchuk, B.; Zheludev, N. I.; Maier, S. A.; Halas, N. J.; Nordlander, P.; Giessen, H.; Chong, C. T. The Fano Resonance in Plasmonic Nanostructures and Metamaterials. *Nat. Mater.* **2010**, *9* (9), 707–715. https://doi.org/10.1038/nmat2810.

(48) Fan, S. Sharp Asymmetric Line Shapes in Side-Coupled Waveguide-Cavity Systems. *Appl. Phys. Lett.* **2002**, *80* (6), 908–910. https://doi.org/10.1063/1.1448174.

(49) Ras, R. H. A.; Schoonheydt, R. A.; Johnston, C. T. Relation between S-Polarized and p-Polarized Internal Reflection Spectra: Application for the Spectral Resolution of Perpendicular Vibrational Modes. *J. Phys. Chem. A* **2007**, *111* (36), 8787–8791. https://doi.org/10.1021/jp073108a.

(50) Cuesta, A.; Couto, A.; Rincón, A.; Pérez, M. C.; López-Cudero, A.; Gutiérrez, C. Potential Dependence of the Saturation CO Coverage of Pt Electrodes: The Origin of the Pre-Peak in CO-Stripping Voltammograms. Part 3: Pt(Poly). *J. Electroanal. Chem.* **2006**, *586* (2), 184–195. https://doi.org/10.1016/j.jelechem.2005.10.006.

(51) López-Cudero, A.; Cuesta, A.; Gutiérrez, C. Potential Dependence of the Saturation CO Coverage of Pt Electrodes: The Origin of the Pre-Peak in CO-Stripping Voltammograms. Part 1: Pt(111). *J. Electroanal. Chem.* **2005**, *579* (1), 1–12. https://doi.org/10.1016/j.jelechem.2005.01.018.

(52) Sun, S.-G.; Zhou, Z.-Y. Surface Processes and Kinetics of CO2 Reduction on Pt(100) Electrodes of Different Surface Structure in Sulfuric Acid Solutions. *Phys. Chem. Chem. Phys.* **2001**, *3* (16), 3277–3283. https://doi.org/10.1039/b100938i.

(53) Rodes, A.; Pastor, E.; Iwasita, T. Structural Effects on CO2 Reduction at Pt Single-Crystal Electrodes: Part 3. Pt(100) and Related Surfaces. *J. Electroanal. Chem.* **1994**, *377* (1), 215–225. https://doi.org/10.1016/0022-0728(94)03424-9.

(54) López-Cudero, A.; Cuesta, Á.; Gutiérrez, C. Potential Dependence of the Saturation CO Coverage of Pt Electrodes: The Origin of the Pre-Peak in CO-Stripping Voltammograms. Part 2: Pt(100). *J. Electroanal. Chem.* **2006**, *586* (2), 204–216. https://doi.org/10.1016/j.jelechem.2005.10.003.

(55) Chang, S.-C.; Weaver, M. J. In-Situ Infrared Spectroscopy of CO Adsorbed at Ordered Pt(110)-Aqueous Interfaces. *Surf. Sci.* **1990**, *230* (1), 222–236. https://doi.org/10.1016/0039-6028(90)90030-C.

(56) Blizanac, B. B.; Lucas, C. A.; Gallagher, M. E.; Arenz, M.; Ross, P. N.; Marković, N. M. Anion Adsorption, CO Oxidation, and Oxygen Reduction Reaction on a Au(100) Surface: The PH Effect. *J. Phys. Chem. B* **2004**, *108* (2), 625–634. https://doi.org/10.1021/jp036483l.

(57) Mayet, N.; Servat, K.; Kokoh, K. B.; Napporn, T. W. Electrochemical Oxidation of Carbon Monoxide on Unsupported Gold Nanospheres in Alkaline Medium. *Electrocatalysis* **2021**, *12* (1), 26–35. https://doi.org/10.1007/s12678-020-00626-7.

(58) Spendelow, J. S.; Goodpaster, J. D.; Kenis, P. J. A.; Wieckowski, A. Mechanism of CO Oxidation on Pt(111) in Alkaline Media. *J. Phys. Chem. B* **2006**, *110* (19), 9545–9555. https://doi.org/10.1021/jp060100c.

(59) Varela, A. S.; Schlaup, C.; Jovanov, Z. P.; Malacrida, P.; Horch, S.; Stephens, I. E. L.; Chorkendorff, I. $CO_2$ Electroreduction on Well-Defined Bimetallic Surfaces: Cu Overlayers on Pt(111) and Pt(211). *J. Phys. Chem. C* **2013**, *117* (40), 20500–20508. https://doi.org/10.1021/jp406913f.

(60) Samjeské, G.; Komatsu, K.; Osawa, M. Dynamics of CO Oxidation on a Polycrystalline Platinum Electrode: A Time-Resolved Infrared Study. *J. Phys. Chem. C* **2009**, *113* (23), 10222–10228. https://doi.org/10.1021/jp900582c.

(61) Katayama, Y.; Giordano, L.; Rao, R. R.; Hwang, J.; Muroyama, H.; Matsui, T.; Eguchi, K.; Shao-Horn, Y. Surface (Electro)Chemistry of $CO_2$ on Pt Surface: An *in Situ* Surface-Enhanced Infrared Absorption Spectroscopy Study. *J. Phys. Chem. C* **2018**, *122* (23), 12341–12349. https://doi.org/10.1021/acs.jpcc.8b03556.

(62) Stamenkovic, V.; Chou, K. C.; Somorjai, G. A.; Ross, P. N.; Markovic, N. M. Vibrational Properties of CO at the Pt(111)−Solution Interface: The Anomalous Stark-Tuning Slope. *J. Phys. Chem. B* **2005**, *109* (2), 678–680. https://doi.org/10.1021/jp044802i.

(63) Dunwell, M.; Wang, J.; Yan, Y.; Xu, B. Surface Enhanced Spectroscopic Investigations of Adsorption of Cations on Electrochemical Interfaces. *Phys. Chem. Chem. Phys.* **2017**, *19* (2), 971–975. https://doi.org/10.1039/C6CP07207K.

(64) Dunwell, M.; Yan, Y.; Xu, B. In Situ Infrared Spectroscopic Investigations of Pyridine-Mediated $CO_2$ Reduction on Pt Electrocatalysts. *ACS Catal.* **2017**, *7* (8), 5410–5419. https://doi.org/10.1021/acscatal.7b01392.

(65) Deshlahra, P.; Conway, J.; Wolf, E. E.; Schneider, W. F. Influence of Dipole–Dipole Interactions on Coverage-Dependent Adsorption: CO and NO on Pt(111). *Langmuir* **2012**, *28* (22), 8408–8417. https://doi.org/10.1021/la300975s.

(66) Jacobse, L.; Rost, M. J.; Koper, M. T. M. Atomic-Scale Identification of the Electrochemical Roughening of Platinum. *ACS Cent. Sci.* **2019**, *5* (12), 1920–1928. https://doi.org/10.1021/acscentsci.9b00782.

(67) Deng, X.; Galli, F.; Koper, M. T. M. In Situ AFM Imaging of Platinum Electrode Surface during Oxidation–Reduction Cycles in Alkaline Electrolyte. *ACS Appl. Energy Mater.* **2020**, *3* (1), 597–602. https://doi.org/10.1021/acsaem.9b01700.

(68) Hammer, B.; Nielsen, O. H.; Nrskov, J. K. Structure Sensitivity in Adsorption: CO Interaction with Stepped and Reconstructed Pt Surfaces. *Catal. Lett.* **1997**, *46* (1), 31–35. https://doi.org/10.1023/A:1019073208575.

(69) Rakić, A. D.; Djurišić, A. B.; Elazar, J. M.; Majewski, M. L. Optical Properties of Metallic Films for Vertical-Cavity Optoelectronic Devices. *Appl. Opt.* **1998**, *37* (22), 5271. https://doi.org/10.1364/AO.37.005271.

(70) Hu, H.; Weber, T.; Bienek, O.; Wester, A.; Hüttenhofer, L.; Sharp, I. D.; Maier, S. A.; Tittl, A.; Cortés, E. Catalytic Metasurfaces Empowered by Bound States in the Continuum. *ACS Nano* **2022**, *16* (8), 13057–13068. https://doi.org/10.1021/acsnano.2c05680.

**Acknowledgments**

We thank Thomas Weber for his help with the temporal-coupled mode theory algorithms and Simon Stork for the platinum and titanium electron beam evaporation.

**Corresponding Author**

Andreas Tittl - Chair in Hybrid Nanosystems, Nanoinstitute Munich, Faculty of Physics, Ludwig-Maximilians-Universität München, Königinstraße 10, 80539 München, Germany; Orcid: https://orcid.org/0000-0003-3191-7164; Email: Andreas.Tittl@physik.uni-muenchen.de

Katharina Krischer - Department of Physics, Technical University of Munich, 85748 Garching, Germany; Email: Krischer@tum.de

**Author Contributions**

The manuscript was written through the contributions of all authors. All authors have given approval to the final version of the manuscript. ‡These authors contributed equally.

**Funding Sources**

This work was funded by the Deutsche Forschungsgemeinschaft (DFG, German Research Foundation) under grant numbers EXC 2089/1–390776260 (Germany′s Excellence Strategy) and TI 1063/1 (Emmy Noether Program), ERC-STG 802989 Catalight, the Bavarian program Solar Energies Go Hybrid (SolTech) and the Center for NanoScience (CeNS). S.A.M. additionally acknowledges the Lee-Lucas Chair in Physics and the EPSRC (EP/W017075/1).

**Notes**

The authors declare no competing financial interest.

# Improved in-situ characterization of electrochemical interfaces using metasurface-driven surface-enhanced infrared absorption spectroscopy

Luca M. Berger [1, ‡], Malo Duportal [2, ‡], Leonardo de Souza Menezes [1,3], Emiliano Cortés [1], Stefan A. Maier [4,5,1], Andreas Tittl [1,*], Katharina Krischer [2,*]

[1] Faculty of Physics, Ludwig-Maximilian-University Munich, 80539 München, Germany
[2] Department of Physics, Technical University of Munich, 85748 Garching, Germany
[3] Departamento de Física, Universidade Federal de Pernambuco, 50670-901 Recife-PE, Brazil
[4] School of Physics and Astronomy, Monash University, Melbourne, Victoria, Australia
[5] Department of Physics, Imperial College London, SW7 2AZ London, United Kingdom

* e-mail: Andreas.Tittl@physik.uni-muenchen.de; krischer@tum.de

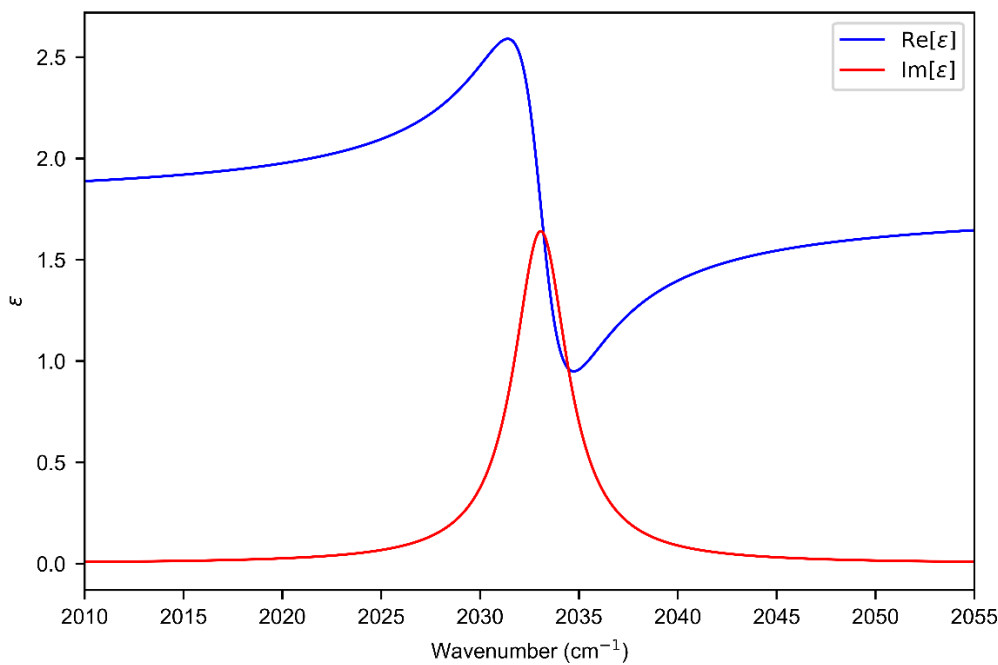

**Figure S1.** The real (blue) and imaginary (red) parts of the permittivity of the artificial material modeled to represent CO. The permittivity of a material suffices to numerically model its interaction with light. Using the Lorentz oscillator model, the real and imaginary parts of the permittivity can be written as $\varepsilon_r = 1 + \frac{\omega_p^2(\omega_0^2 - \omega^2)}{(\omega_0^2 - \omega^2)^2 + \gamma^2\omega^2}$ and $\varepsilon_i = \frac{\gamma\omega_p^2\omega}{(\omega_0^2 - \omega^2)^2 + \gamma^2\omega^2}$, respectively[1]. Here, we modified the baseline of $\varepsilon_r$ to be around $1.33^2$ which is the refractive index squared used for the surrounding medium (water) due to a negligible shift in the refractive index due to $CO^2$. Here, $\omega_p$ is analogous to the plasma frequency in the Drude-Sommerfeld model, $\omega_0$ is the frequency of the absorption band, $\omega$ is the frequency, and $\gamma$ is the damping constant. To model the linear vibrational mode of CO at 2033 cm$^{-1}$ the parameters used were $\omega_p^2 = 1 \times 10^{24}$ s$^{-2}$, $\omega_0 = 60.95 \times 10^{12}$ s$^{-1}$ (corresponding to 2033 cm$^{-1}$), and $\gamma = 1 \times 10^{11}$ s$^{-1}$.

## References

(1)   Novotny, L.; Hecht, B. *Principles of Nano-Optics*; Cambridge University Press, 2012.

(2)   Harvey, A. H.; Kaplan, S. G.; Burnett, J. H. Effect of Dissolved Air on the Density and Refractive Index of Water. *Int. J. Thermophys.* **2005**, *26* (5), 1495–1514. https://doi.org/10.1007/s10765-005-8099-0.